\documentclass{PoS}
\usepackage{amssymb,amsmath}
\usepackage{epsfig,psfrag}

\newcommand{\defphi}{\phi}
\newcommand{\eps}{\epsilon}

\title{The $n_{f}$ terms of the three-loop cusp anomalous dimension in QCD}
\ShortTitle{On the QCD cusp anomalous dimension}

\author{{Andrey Grozin}\\
        Budker Institute of Nuclear Physics, Novosibirsk, Russia\\
        Novosibirsk State University, Novosibirsk, Russia\\
        E-mail: \email{A.G.Grozin@inp.nsk.su}}

\author{\speaker{Johannes M. Henn}\\
        Institute for Advanced Study, Princeton, NJ 08540, USA\\
        E-mail: \email{jmhenn@ias.edu}}

\author{{Gregory P. Korchemsky}\thanks{Preprint number: IPhT-T14-093.}
\\
        Institut de Physique Th\'eorique, CEA Saclay, 91191 Gif-sur-Yvette Cedex, France\\
        E-mail: \email{Gregory.Korchemsky@cea.fr}}

\author{{Peter Marquard}\thanks{Preprint numbers: DESY 14-113, LPN14-080, SFB/CPP-14-33.}
\\
        Deutsches Elektronen-Synchrotron, DESY, Platanenallee 6, D15738 Zeuthen, Germany\\
        E-mail: \email{peter.marquard@desy.de}}

\abstract{In this talk we present the result for the $n_f$ dependent piece of the three-loop cusp anomalous
dimension in QCD. Remarkably, it is parametrized by the same simple functions appearing in analogous anomalous dimensions in ${\mathcal N}=4$ SYM
at one and two loops. 
We also compute all required master integrals using a recently proposed refinement of the differential equation method.
The analytic results are expressed in terms of harmonic polylogarithms of uniform weight. 
}

\FullConference{ Loops and Legs in Quantum Field Theory - LL 2014,\\
		 27 April - 2 May 2014 \\
		 Weimar, Germany }

\begin{document}

\section{Introduction}

{The cusp anomalous dimension is an ubiquitous quantity in gauge theories. It governs
the dependence of the cusped Wilson loop on the ultraviolet cut-off  \cite{PolyakovCusp} and appears in many physical
quantities,  e.g. it controls the infrared asymptotics of 
scattering amplitudes and form factors involving massive particles \cite{Korchemsky:1991zp,hqetreviews}. The two-loop result for this fundamental quantity has been known for more than 25 years 
\cite{twoloop}.
Here we report on a calculation of the $n_f$-dependent contribution to the cusp anomalous dimension  in QCD at three 
loops.}

\section{Overview of results in $\mathcal N=4$ SYM and QCD}

Recent years have seen a lot of progress in understanding the cusp anomalous
dimension in $\mathcal N=4$ supersymmetric Yang-Mills (SYM), where perturbative results 
are available to three and four loops, including part of the non-planar corrections which first 
appear at four loops \cite{3and4loops}.\footnote{Obviously, the perturbative regime is most relevant for the comparison with QCD.
However, we would also like to mention that results are available at strong coupling \cite{adscft}, via
the AdS/CFT correspondence. Moreover, exact results are known in the small angle regime \cite{smallangle},
and there is an approach based on integrability, cf. \cite{integrableWL} and references therein.
The cusp anomalous dimension can also be obtained from the Regge limit of certain massive scattering amplitudes \cite{Henn:2010bk}.}
The cusp anomalous dimension takes a particularly simple form in $\mathcal N=4$ SYM and 
it can be organized according to the transcendental weight of contributing functions. In this section,
we review these results in order to compare them to the QCD answer.

The most natural Wilson loop operator to consider in $\mathcal N=4$ SYM has an additional coupling to
scalars \cite{adscft} depending on a unit vector $n^I$ in the internal $S^5$ space, $(n^I)^2=1$, and
an auxiliary parameter $\sigma$
\begin{align}\label{W-sigma}
W_\sigma =\langle 0|\,{\rm tr} \bigg[ P \exp\left(i \oint_C dx\cdot A(x) + \sigma\oint_C  d |x| \,  n^I \,\phi_I(x)  \right)\bigg] |0\rangle \,.
\end{align}
For $\sigma=1$, the Wilson loop $W_{\sigma=1}$ locally preserves supersymmetry whereas for $\sigma=0$ it coincides
with the conventional Wilson loop with only coupling to gluons as in QCD.\footnote{It is not known at present whether integrability extends to this case.} 
We will refer to the $\sigma=1$ and $\sigma=0$ cases as the supersymmetric and bosonic Wilson loop, respectively.

\begin{figure}
\centerline{\psfrag{v1}[cc][cc]{$v_1$} \psfrag{v2}[cc][cc]{$v_2$} 
\psfrag{phi}[cc][cc]{$\phi$}
  \includegraphics[width = 60mm]{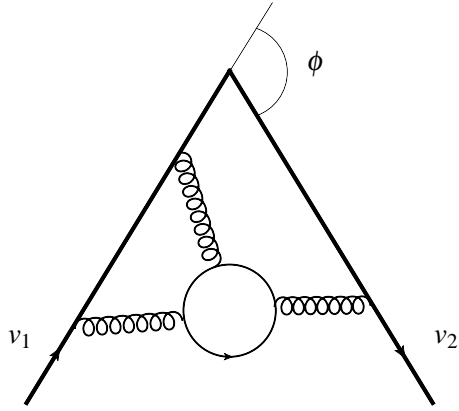}}
  \caption{Sample Feynman diagram producing an $n_f$ dependent contribution to 
the three-loop cusp anomalous dimension in QCD. Thick
  lines denote two semi-infinite segments forming a cusp of angle 
$\phi$. Wavy
  lines stand for gauge fields and the thin circle for a quark loop.}
  \label{figurenf}
\end{figure}

{To compute the cusp anomalous dimension, we consider an integration contour $C$ formed by two segments 
along space-like directions $v_{1}^\mu$ and $v_{2}^\mu$ (with $v_1^2= v_2^2 =1$), with cusp angle 
${\cos \phi = v_{1} \cdot v_{2}}$ (cf. Fig.~\ref{figurenf}). In addition, we take the vectors $n_1^I$ and $n_2^I$ to be constant along the segments 
except the cusp point where they form an additional internal angle $\cos\theta=n_1^I n_2^I$.
The cusp anomalous dimension depends on the cusp angles $\phi$ and $\theta$. 
It turns out to be convenient to introduce complex variables
\begin{align}
x=e^{i \phi}\,,\qquad\qquad  \xi = (\cos \theta - \cos \phi )/(i \sin \phi)
\end{align}
}
The dependence  { of the cusp anomalous dimension} on $\xi$ is polynomial.
For simplicity of notation, let us set $\theta = \pi/2$ from now on, i.e. $\xi = {(1+x^2)}/{(1-x^2)}$.

The two-loop results for the Wilson loop operators $W_{\sigma=1}$ and $W_{\sigma=0}$ in $\mathcal N=4$ SYM are\footnote{The supersymmetric results quoted here are valid in the DRED scheme,
while formulas in QCD will be given in the $\overline{\rm MS}$ scheme. See Appendix A of ref. \cite{Belitsky:2003ys} for a discussion of the scheme conversion up to two loops.}
\begin{align}\label{resultN4SYMsusy}
\Gamma_{\rm SYM}^{\rm susy \;WL} =&  \,
a\,
A^{(1)}(\phi)
+a^2 \, A^{(2)}(\phi) \,, \\
\Gamma_{\rm SYM}^{\rm bosonic \;WL}  =& \, a\,
\left[ A^{(1)}(\phi) -  A^{(1)}(0) \right] +   a^2\, \left[ A^{(2)}(\phi) - A^{(2)}(0) + B^{(2)}(\phi) - B^{(2)}(0) \right] \,,\label{resultN4SYMbosonic}
\end{align}
where $a = g^2 N/(8 \pi^2)$ is the 't Hooft coupling and
\begin{align} \notag \label{A-B}
A^{(1)}(\phi) =& -  \xi \,  \log x\,, \\[2mm] 
B^{(2)}(\phi)  =&  2 \zeta_2 +  \log^2 x    - \xi \left[ \zeta_2 + \log^2 x + 2\, {\rm Li}_{1}(x^2)  \log x  - {\rm Li}_{2}(x^2) \right] \,, \\
A^{(2)}(\phi) =& \;\;  \xi \, \left[ 2\zeta_2 \log x+ \frac{1}{3} \log^3 x \right]   -  \xi^2\, \left[ \zeta_3 + \zeta_2 \log x+\frac{1}{3} \log^3 x + {\rm Li}_{2}(x^2) \log x  - {\rm Li}_{3}(x^2)  \right]  \,.
\notag
\end{align}
Eq. (\ref{resultN4SYMsusy}) is due to the last ref. in \cite{twoloop},
while to the best of our knowledge eq. (\ref{resultN4SYMbosonic}) is new. 
Note that although each of the functions (\ref{A-B}) has uniform weight $1$,$2$ and $3$, respectively,
they produce a `weight drop' contribution
when evaluated at zero angle, $A^{(1)}(0) = 1$,  $B^{(2)}(0) =-2+2\zeta_2$, and $A^{(2)}(0) = 1- 2\zeta_2$.

Interestingly, the cusp anomalous dimension for the bosonic Wilson loop in $\mathcal N=4$ SYM differs 
only slightly from the supersymmetric one.
Moreover, the function $B^{(2)}$ is related to a derivative of $A^{(2)}$, if one considers $\xi$ and $x$
as independent variables,
\begin{align}
B^{(2)} = {1\over \xi} {\partial \over \partial \log x} A^{(2)}
\,.
\end{align}
Using relations (\ref{A-B}), we can rewrite the known two-loop result for the QCD cusp anomalous dimension in a new way,
 in terms of the simple functions encountered in $\mathcal N=4$ SYM,
\begin{align}
\Gamma_{\rm QCD}^{(1)} ={}& C_{F} \left[ A^{(1)}(\phi) - A^{(1)}(0) \right] \,, \\
\Gamma_{\rm QCD}^{(2)} ={}& 
\frac{1}{2}
 C_{F} C_{A} \left[ A^{(2)}(\phi) - A^{(2)}(0) + B^{(2)}(\phi) - B^{(2)}(0) \right] \nonumber \\ & +  \left(\frac{67}{36} C_{F} C_{A} - \frac{5}{9} 
C_{F} T_{f} n_{f}  \right)   \left[ A^{(1)}(\phi) - A^{(1)}(0) \right] \,,
\end{align}
where the expansion parameter is ${\alpha_s}/\pi$, $C_F$ and $C_A$ are quadratic Casimirs of the $SU(N)$ gauge group in the fundamental and adjoint representation, respectively,   $n_f$ is the number of quark flavours and $T_f=1/2$.

\section{Uniform weight functions and computation of the master integrals}

Why should uniform weight functions play such an important role for the cusp anomalous dimension? In fact,
the perturbative expansion of a cusped Wilson loop (\ref{W-sigma}) gives rise to distinct Feynman integrals which are 
already very close to the definition of iterated integrals \cite{Chen:1977oja}. In the third reference of \cite{3and4loops}, 
this observation was used to give an algorithm for computing any Wilson line integral with an arbitrary number of 
propagator exchanges (but no internal vertices).\footnote{A different computation of some of these integrals is discussed in \cite{Gardi}.}
For the full computation we require a larger class of integrals that includes graphs with interaction vertices.
A method which exposes the weight properties of such integrals was proposed in \cite{Henn:2013pwa},
and we used it for our computation.

Since the three-loop cusp anomalous dimension does not receive nonplanar corrections, it can be expressed in 
terms of  planar integrals only. 
We choose to perform the calculation in momentum space, using
the heavy quark effective theory framework [3]. The integrals
can all be parametrized as (with $D=4-2\eps$)
\begin{align}\label{G}
 G_{a_1, \ldots, a_{12}} =& e^{3 \eps \gamma_{\rm E}} \int \frac{d^{D}k_{1}d^{D}k_{2}d^{D}k_{3}}{(i \pi^{D/2})^3} 
 (-2k_1\cdot v_1 +1)^{-a_1}(-2k_2\cdot v_1 +1)^{-a_2}(-2k_3\cdot v_1 +1)^{-a_3} \nonumber \\
 & \quad \times (-2k_1\cdot v_2 +1)^{-a_4}(-2k_2\cdot v_2 +1)^{-a_5}(-2k_3\cdot v_2 +1)^{-a_6}
 (-k_1^2)^{-a_7}\nonumber \\
 & \quad \times (-(k_1-k_2)^2)^{-a_8}(-(k_2-k_3)^2)^{-a_9}(-(k_1-k_3)^2)^{-a_{10}}(-k_2^2)^{-a_{11}}(-k_3^2)^{-a_{12}} \,,
\end{align}
for certain choices  of positive/negative integers $a_{i}$. Applying the
integral reduction algorithms \cite{intreduction}, we found that 71 master integrals are required in total.\footnote{A subset of these integrals that reduce to a one-loop triangle with $\eps$-dependent indices was computed in ref. \cite{Grozin:2011rs}.}
We then used the method proposed in ref. \cite{Henn:2013pwa} to choose a convenient basis for the latter, denoted by $\vec{f}(x,\eps)$.
A distinguished feature of this basis is that the $\vec{f}(x,\eps)$ satisfy the differential equations of the form ($D=4-2\eps$)
\begin{align}\label{DEhqet}
\partial_{x} \, \vec{f}(x,\eps) =  \eps \, \left[ \frac{a}{x} + \frac{b}{x+1} + \frac{c}{x-1} \right] \vec{f}(x,\eps)\,,
\end{align}
with constant ($x-$ and $\eps-$independent) matrices $a,b,c$.
We see that eq. (\ref{DEhqet}) has four regular singular points, $0,1,-1,\infty$. Due to the $x \leftrightarrow {1}/{x}$ symmetry of the definition $2 \cos \phi =x+1/x$, only the first
three are independent. They correspond, in turn, to the light-like limit (infinite angle), to the zero angle limit, and to the threshold limit. 
See ref.  \cite{3and4loops} for further discussion of these limits.
\begin{figure}
\centerline{\psfrag{v1}[cc][cc]{$v_1$} \psfrag{v2}[cc][cc]{$v_2$} 
\psfrag{k1}[cc][cc]{$k_1$} \psfrag{k2}[cc][cc]{$k_2$}
\psfrag{k12}[cc][cc]{$k_1-k_2$}\psfrag{k23}[cc][cc]{$k_2-k_3$}
  \includegraphics[width = 60mm]{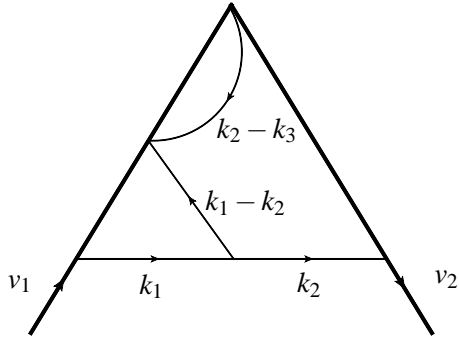}}
  \caption{Diagrammatical representation of the basis integral $f_{44}$ discussed in the main text. Thin 
lines denote scalar propagators. One of the propagators is doubled, and a normalization factor is not shown in the figure.}
\end{figure}

Solving (\ref{DEhqet}) we use boundary conditions for $\vec{f}(x,\eps)$ at $x=1$. 
All $\vec{f}(1,\eps)$ except one can be easily obtained from consistency conditions,
i.e. absence of unphysical singularities, and the remaining constant can be found by comparing to refs. \cite{Chetyrkin:2003vi}.
If follows immediately from (\ref{DEhqet}) that the solution for $\vec{f}$ in the from of $\eps-$expansion can be written in terms of harmonic 
polylogarithms \cite{Remiddi:1999ew}.
In this way, we obtained an analytic answer in terms of uniform weight functions for all integrals required.
As an example, we consider one of the master integrals 
\begin{align}\notag
f_{44}= \eps^5 \frac{1-x^2}{x} G_{1,0,1,0,1,0,1,1,2,0,1,0} = \eps^4 \Big[ -\frac{1}{6} \pi^2 H_{0,0}(x) -\frac{2}{3}\pi^2 H_{1,0}(x) -4 H_{0,-1,0,0}(x) 
\\
+2 H_{0,0,-1,0}(x) +2 H_{0,1,0,0}(x) 
-4 H_{1,0,0,0}(x) +4 \zeta_{3} H_{0}(x) - \frac{17 \pi^4}{360} \Big] + {\mathcal O}(\eps^5) \,. \label{basisint44}
\end{align}
We performed numerical checks on all integrals using FIESTA \cite{Smirnov:2008py},
and analytically reproduced results for three-loop integrals known from the ${\mathcal N}=4$ SYM computation \cite{3and4loops}.
More details will be discussed elsewhere.

\section{Three-loop cusp anomalous dimension in QCD}

To compute the cusp anomalous dimension, we started with the definition of the bosonic Wilson loop (\ref{W-sigma}) in QCD,
and generated all Feynman diagrams contributing to $W_{\sigma=0}$ up to three loops, in an arbitrary covariant gauge.
This was done with the help of the computer programs QGRAF and FORM \cite{Nogueira:1991ex}.
The $n_f-$dependent contribution only comes from diagrams with quarks propagating inside the loops. They can be evaluated
within dimensional regularization using standard methods, details will be given elsewhere. 

The cusp anomalous dimension $\Gamma$ can be extracted from the divergent part of the one-particle irreducible 
vertex function $V(\phi)$ of the heavy-to-heavy current \cite{hqetreviews} (e.g., the diagram in Fig.~\ref{figurenf} without the external leg propagators),
 \begin{align}\label{defZJ}
\log V(\defphi) - \log V(0)   = \log Z + {\mathcal{O}(\eps^0)}\,,\qquad  \Gamma=\frac{d \log Z }{d\log\mu} \,.
\end{align}
In the $\overline{\rm MS}$ scheme,  the renormalization   $Z-$factor has the following structure
\begin{align}\label{structuredivergences}
\log Z =& 
- \frac{1}{2 \eps} \left(\frac{\alpha_{s}}{\pi} \right)  \Gamma^{(1)} 
+  \left(\frac{\alpha_{s}}{\pi} \right)^2  \left[  \frac{\beta_{0}}{16 \eps^2}  \Gamma^{(1)} - \frac{1}{4 \eps} \Gamma^{(2)}  \right] 
\nonumber \\ &\quad\quad 
 + \left(\frac{\alpha_{s}}{\pi} \right)^3  \left[ - \frac{\beta_0^2 \Gamma^{(1)}}{96 \eps^3} + \frac{\beta_1 \Gamma^{(1)} + 4 \beta_0 \Gamma^{(2)} }{96 \eps^2}  - \frac{\Gamma^{(3)}}{6 \eps}  \right] \,,
\end{align}
where the $\mu-$dependence only enters through the renormalised coupling constant  \cite{vanRitbergen:1997va}, $\frac{d}{d\log\mu} \left(\frac{\alpha_{s}}{4 \pi}\right) = -2 \eps  \left(\frac{\alpha_{s}}{4 \pi}\right)  - 2 \beta(\alpha_s) $.
%
As a non-trivial check of our calculation we verified that eq.~(\ref{structuredivergences}) indeed reproduces the pole structure of $\log V(\phi)$
at three loops.
 
At three loops, the cusp anomalous dimension has the following form by virtue of non-Abelian exponentiation,
\begin{align}\label{gen}
\Gamma_{\rm QCD}^{(3)} = c_1\, C_{F} C_{A}^2 + c_{2}\, C_F (T_f n_f)^2 + c_3\, C_{F}^2 T_f n_f + c_4\, C_{F} C_{A} T_{f} n_{f} \,.
\end{align}
For the $n_f$ dependent terms, we obtained the following results,
\begin{align} \notag  
c_{2} = {}& -\frac{1}{27} \left[ A^{(1)}(\phi) - A^{(1)}(0) \right] \,,\\ 
c_{3} = {} & \left( \zeta_3 - \frac{55}{48} \right)  \left[ A^{(1)}(\phi) - A^{(1)}(0) \right] \,, \label{resultc4} \\
c_{4} = {} & -\frac{5}{9}
 \left[ A^{(2)}(\phi) - A^{(2)}(0)+B^{(2)}(\phi) - B^{(2)}(0) \right]  - \frac{1}{6} \left( 7 \zeta_{3} + \frac{209}{36} \right) \left[ A^{(1)}(\phi) - A^{(1)}(0) \right]\notag
\,. 
\end{align} 
with the functions $A^{(1)}$, $A^{(2)}$ and $B^{(2)}$ given in eq. (\ref{A-B}).

The following comments are in order. The leading 
$n_f^2$
term in  (\ref{gen}) is in agreement with the known result \cite{Beneke:1995pq}.
The expressions for the coefficients $c_3$ and $c_4$ in the subleading 
$n_{f}$
terms are new ($c_3$ can be obtained by generalizing the method of the last 
ref. of  \cite{hqetreviews}).

As yet another check of our result, 
we can take the light-like limit of  (\ref{gen}), where one expects \cite{limitx0} the behavior
$\lim_{x\to0} \Gamma \to  K(\alpha_s) \log (1/x) $, with $K$ at three loops computed in refs. \cite{lightlike}.
Again, we observed a perfect agreement for the $n_f$ dependent terms.

It is remarkable that despite the relative complexity of the Feynman integrals (\ref{G}), 
the final expressions (\ref{resultc4}) are surprisingly simple!
Moreover, they are expressed in terms of the same functions that appear in the $\mathcal N=4$ SYM answer. 
It will be interesting to see whether this is also the case for the $C_F C_A^2$ term. 
This calculation is work in progress.

\section{Discussion}

The simplicity of eqs. (\ref{resultc4}) suggests that there should be a simpler way of arriving at
these results. Ignoring technical details such as the intrinsic renormalization of the Lagrangian and the associated $\beta$ function,
morally speaking there should be a way of organizing the calculation in terms of manifestly finite integrals in four dimensions, 
as in ref. \cite{Caron-Huot:2014lda}. 
This would very likely require only a (simpler) subset of functions as compared to the calculation in $D=4-2\eps$ dimensions.

A related comment is that when  computing integrals via differential equations,
usually one proceeds in a ``bottom-up'' approach: one starts with the integrals with few propagators, e.g. a tadpole integral,
when proceeds with bubbles, and so on. 
Let us now imagine a scenario where, through some means, one knows the answer for $\mathcal N=4$ SYM. 
The integrals required for $\mathcal N=4$ SYM are typically the ones with maximal number of propagators, thanks to its good
ultraviolet properties. In the traditional approach, one arrives at them only at the very end, and therefore they obviously contain a lot
of information. Given this, it is interesting to ask whether one can use this information in a ``top-down'' approach, and
how many of the master integrals required for QCD are determined by it.

\acknowledgments
We wish to thank R.~N.~Lee for his help in using LiteRed, and K.~G.~Chetyrkin, M.~Steinhauser and V.~Smirnov 
for helpful conversations. We are also grateful to several institutes that have provided hospitality to some of us 
during scientific visits in the course of this work, namely IPhT Saclay, the Karlsruhe Institute of Technology (KIT), 
and the Institute for Advanced Study, Princeton.
A.G.'s work was supported by RFBR grant 12-02-00106-a 
and by the Russian Ministry of Education and Science.
J.M.H. is supported in part by
the DOE grant DE-SC0009988 and by the Marvin L. Goldberger fund.
G.P.K. is supported in part by 
the French National Agency for Research (ANR) under contract StrongInt
(BLANC-SIMI-4-2011).
P.M. was supported in part by the DFG through the SFB/TR 9 ``Computational
Particle Physics'' and the EU Networks LHCPHENOnet PITN-GA-2010-264564
and HIGGSTOOLS PITN-GA-2012-316704.

\end{document}